\documentclass[aps,prl,reprint,subcriptaddress]{revtex4-1}
\usepackage{graphicx}
\usepackage{amsmath}
\usepackage{amssymb}
\usepackage{amsfonts}
\usepackage{dcolumn}

\begin{document}
\title{Can Anomalous Amplification be Attained without Postselection?}

\author{Juli\'{a}n Mart\'{i}nez-Rinc\'{o}n$^{1,3}$}
\email{jmarti41@ur.rochester.edu}
\author{Wei-Tao Liu$^{2,1,3}$}
\author{Gerardo I. Viza$^{1,3}$}
\author{John C. Howell$^{1,3,4,5}$}

\address{$^1$Department of Physics and Astronomy, University of Rochester, Rochester, New York 14627, USA}
\address{$^2$College of Science, National University of Defense Technology, Changsha 410073, P.R.China}
\address{$^3$Center for Coherence and Quantum Optics, University of Rochester, Rochester, New York 14627, USA}
\address{$^4$Institute of Optics, University of Rochester, Rochester, New York 14627, USA}
\address{$^5$Institute for Quantum Studies, Chapman University, 1 University Drive, Orange, California 92866, USA}


\date{\today}

\begin{abstract}
We present a parameter estimation technique based on performing joint measurements of a weak interaction away from the weak-value-amplification approximation. Two detectors are used to collect full statistics of the correlations between two weakly entangled degrees of freedom. Without discarding of data, the protocol resembles the anomalous amplification of an imaginary-weak-value-like response. The amplification is induced in the difference signal of both detectors allowing robustness to different sources of technical noise, and offering in addition the advantages of balanced signals for precision metrology. All of the Fisher information about the parameter of interest is collected. A tunable phase controls the strength of the amplification response. We experimentally demonstrate the proposed technique by  measuring polarization rotations in a linearly polarized laser pulse. 
We show that in the presence of technical noise the effective sensitivity and precision of a split detector is increased when compared to a conventional continuous-wave balanced detection technique. 
\end{abstract}

\maketitle 

\paragraph{Introduction} 
Anomalous amplification~\cite{AAV} has been shown to be advantageous for precision metrology. 
Such an amplification provides a way to increase a signal while decreasing~\footnote{The decrease of the noise floor is possible when the dominant source of technical noise is dependent on the number of detected events. For such a case, discarding of data plays an important role. See details in Ref.~\cite{PRX,NoiseExp}.} or retaining the technical-noise floor~\cite{PRX,NoiseExp}. As a result, the sensitivity and precision of measurements limited by technical noise can be effectively improved, facilitating the saturation of the standard quantum limit. Anomalous amplification was first proposed for metrology with the introduction of the Weak Value (WV) of an observable~\cite{AAV,Duck}, and parameter estimation protocols defined after it are usually known as Weak-Value-Amplification (WVA) techniques. The WV of an observable is obtained by post-selecting the state of a system after a weak interaction with a meter system. 
In WVA, such measurements in the system induce a discarding of data counts in the measurements of the meter. In addition to the notion that the state of the system is post-selected after the weak interaction, we consider post-selection as the process of selecting and processing desired events, which, for WVA, results in discarding data in the meter.
Due to the interference of the pre- and post-selection states of the system the WV can take large complex values outside the eigenvalue spectrum of the observable, which defines the anomalous amplification in WVA. Discussion about the quantum interpretation of such a phenomenon can be found in Refs.~\cite{jozsa,ProofContextuality,Interference,WVQuantum}. Many recent applications of WVA for metrology have been done in classical optics, where the interference can be understood using standard wave mechanics~\cite{John,WVClassical}.   

Strong postselection is necessary for anomalous amplification in WVA techniques, but discarding data counts has been the target of criticism, and even considered \textquotedblleft harmful\textquotedblright\,for metrology~\cite{Ferrie1, Combes, StateOfPlay}. However, it has been shown, theoretically and experimentally, that the statistical information collected by the measurements is insignificantly reduced because the amplified signal can compensate for the reduced detection flux resulting from postselection~\cite{Starling,Velocimetry,Nishizawa,PRX,Lee,NoiseExp,Gabriel}. Such a result is possible under an almost orthogonal pre- and post-selection procedure in the system allowing one to collect \textit{nearly all} of the Fisher information using only a small subensemble of measurement counts in the meter~\cite{PRX}. For example, the shot-noise limit defined by the \textit{input} number of photons used in measurements of the small velocity of one of the mirrors of a Michelson interferometer can be reached using WVA~\cite{Velocimetry}. Moreover, it was recently shown that by postselecting only 1$\%$ of the photons, when measuring small optical deflections constrained to intrinsic electronic detector noise, 99$\%$ of the total available Fisher information can be recovered~\cite{NoiseExp}.


Signal amplification while avoiding detector saturation sparked interest in WVA as a precision metrological technique several years ago~\cite{Hosten,Dixon,Brunner}, and was recently shown to be essential for the technical-noise mitigation advantages~\cite{PRX,NoiseExp}. We will introduce the possibility of inducing anomalous amplification without the need of discarding data and without any loss of Fisher information. 





Str\"{u}bi and Bruder~\cite{Strubi} recently proposed a precision measurement technique for measuring time delays of light by carrying out a full measurement of the two weakly correlated degrees of freedom: frequency and polarization. They concluded that even for low-resolution detectors, the scheme is robust against systematic errors and fluctuations in alignments of the experimental setup. We show that by subtracting the readouts of the two detectors in such a measurement, a WVA-like response is obtained in the difference signal. This anomalous amplification behaviour clarifies and extends the results reported in Ref.~\cite{Strubi}, since besides having the amplification response typical of the WVA approach (without discarding data), our protocol adds the benefits of balanced detection for precision metrology. We show that this technique allows us to recover \textit{all} of the Fisher information of the estimated parameter. 
In addition, it permits the removal of systematic error in the measurement of the shift in the difference signal by tracking the unshifted sum signal as well. 

\begin{figure}
\centering
\includegraphics[width=0.45\textwidth]{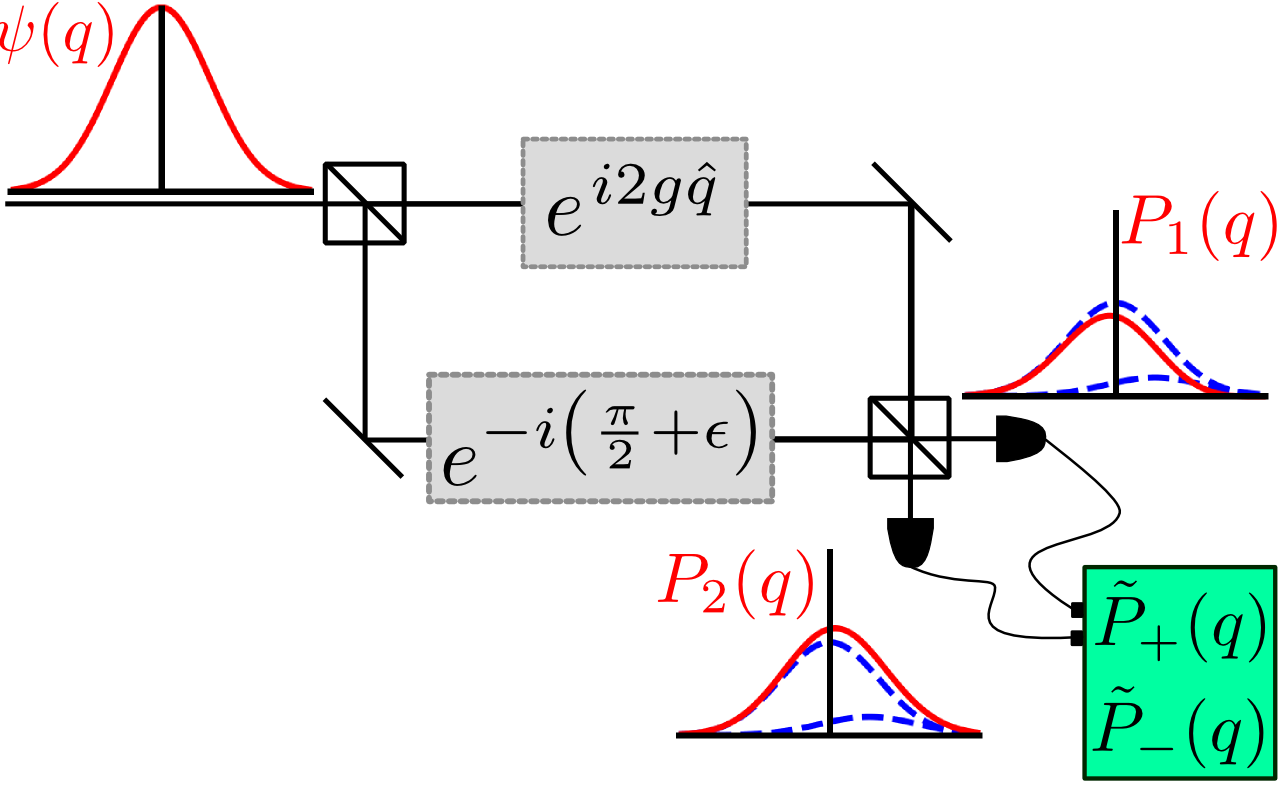}
\caption{\textit{Almost}-balanced weak values detection technique for measuring a small parameter $g$ in an optical setup. The states $|0\rangle$ and $|1\rangle$ correspond to the two paths in the interferometer, a small unbalacing phase $\epsilon$ controls the interference, and estimation of $g$ is obtained from the sum 
and difference 
of the two measured distributions. The red solid lines represent the directly measured distributions $P_1$ and $P_2$, and the blue dashed lines are the distribution components as in Eq.~(\ref{distributions}).}
\label{fig:cartoon}
\end{figure}

\paragraph{Theoretical Framework}
Following the formalism as in WVA, we describe the unitary evolution of a two-party system as $U=\exp{(-ig\,\hat{q}\otimes\hat{A})}$, where $\hat{A}$ is a binary degree of freedom (qubit) controlling the encoding of the information about the interaction parameter $g$ in the continuous (meter) observable $\hat{q}$~\footnote{Such an interaction can be given, for example, by an interaction Hamiltonian of the form $H_{int}=\hbar\tilde{g}(t)\,\hat{q}\otimes\hat{A}$, where $\tilde{g}(t)$ is the instantaneous interaction parameter between the two parties, and $g=\int_0^T\tilde{g}(t)dt$ is the averaged measured parameter, where $T$ is the interaction time.}. In contrast to WVA, where a small set of measurements for $\hat{q}$ are taken into account due to postselection, the operator $\hat{q}$ after the interaction is always measured and conditioned to one of two detectors. 
The proposed procedure is done by preparing the initial global state as the product state $|\Psi_{in}\rangle\otimes\psi(q)$, with $|\Psi_{in}\rangle=(|0\rangle+|1\rangle)/\sqrt{2}$, and by tracking $q$ conditioned to the measurement basis $|\Psi_{1,2}\rangle=(|0\rangle\pm ie^{i\epsilon}|1\rangle)/\sqrt{2}$ on the qubit/system.
Here $|0\rangle$ and $|1\rangle$ are the eigenvectors of $\hat{A}$, where $\hat{A}=|1\rangle\langle1|-|0\rangle\langle0|$, and projections to the two detectors are labelled by 1 and 2. The small phase $\epsilon$ defines the measuring basis on the equatorial plane of the Bloch sphere, where the initial prepared state for the qubit $|\Psi_{in}\rangle$ also lies. The probability distributions measured on the detectors take the form 
\begin{eqnarray}\label{distributions}
P_{1,2}(q;g)&=&\left|\langle \Psi_{1,2}|U|\Psi_{in}\rangle\,\psi(q)\right|^2\nonumber\\&=&\frac{1}{2}\left[1\mp\sin(\epsilon+2gq)\right]P(q),
\end{eqnarray}
where $P(q)=\left|\psi(q)\right|^2$. We will use a Gaussian state $\psi(q)$ with variance $\sigma^2$ for the preparation in $q$. 
As an example, Fig.~\ref{fig:cartoon} shows a schematic of the measuring technique for an optical setup, where the observable $\hat{A}$ is represented as the which-path degree of freedom in an interferometer, and the phase $\epsilon$ controls the interference. 

Under the assumption of a weak interaction, i.e. $2g\sigma\ll\mbox{min}\left\{1,\tan{\epsilon}\right\}$, we can express
\begin{equation}\label{approximation}
\sin(\epsilon+2gq)P(q)\approx\sin(\epsilon)\,P\left(q-2g\sigma^2\cot\epsilon\right).
\end{equation}

The peak value of the distribution of Eq.~(\ref{approximation}) is smaller by a factor $\sin\epsilon$ and the position of the peak is shifted by an amount $\delta q=2g\sigma^2\cot\epsilon$ with respect to $P(q)$. The sum and difference of the distributions take the form~\footnote{We use $\sim$ to refer to post-processing probability distributions. The distributions $P_1$ and $P_2$ are directly obtained from measurements.}
\begin{eqnarray}
\tilde{P}_+(q)&=&P_1+P_2=P(q),\label{sumanddiff1}\\
\tilde{P}_-(q;g)&=&P_2-P_1\approx\sin(\epsilon)\,P\left(q-2g\sigma^2\cot\epsilon\right).\label{sumanddiff2}
\end{eqnarray}   
After a large number $N$ of independent measurements of $q$, $q^{(1)}=q_1,...,q_{N_1}$ on detector 1 and $q^{(2)}=q_1,...,q_{N_2}$ on detector 2, the sum distribution~(\ref{sumanddiff1}) reproduces the quantum probability distribution of the input state, and the difference distribution~(\ref{sumanddiff2}) has a shifted attenuated peak similar to WVA. Note that the measured shift is in the difference probability distribution and not in the wave-function $\psi(q)$ itself, as it is the case in WVA. In fact, the weak values for the measurements are given by $A_w^{1,2}=\langle\Psi_{1,2}|\hat{A}|\Psi_{in}\rangle/\langle\Psi_{1,2}|\Psi_{in}\rangle=\mp i\cos\epsilon/\left(1\mp\sin\epsilon\right) \sim\mp i(1\pm\epsilon)$ for $\epsilon\ll1$, and no (anomalous) large weak value is induced. Estimations of averaged values for $\epsilon$ and $g$ under the weak interaction approximation can be obtained as 
\begin{eqnarray}
\epsilon&=&\sin^{-1}\left(\frac{N_2-N_1}{N_1+N_2}\right),\label{epsilon}\\
g&=&\frac{\left[\langle q\rangle_--\langle q\rangle_+\right]\tan\epsilon}{2\sigma^2}\label{g},
\end{eqnarray}
where $\langle q\rangle_{\pm}=\left(\sum_{i=1}^{N_2} q_i^{(2)} \pm \sum_{i=1}^{N_1} q_i^{(1)} \right)/\left(N_2\pm N_1\right)$, and $\sigma^2$ is the measured variance of $\tilde{P}_+(q)$. By making $\epsilon\ll1$ and preparing a large variance input state, a large shift is induced and small values of $g$ resolved. This behaviour is similar to the amplification of the WVA technique with an imaginary weak value~\cite{jozsa}, and the quadratic response of the WVA postselection probability with respect to $\epsilon$, $\sin^2(\epsilon/2)\sim\epsilon^2/4$, is replaced with a (larger) linear response of the difference signal, $\sin\epsilon\sim\epsilon$ in Eq.~(\ref{sumanddiff2}).

The maximum amount of information about the parameter $g$ that can be extracted from the measurements is given by adding the Fisher information for both detectors~\cite{Supplemental}, 
\begin{equation}\label{Fisher}
F_g=F_1+F_2=4N\sigma^2
\end{equation}
where $F_i=N_i\int_{-\infty}^{\infty} \,\frac{1}{P_i}\left(\frac{\partial P_i}{\partial g}\right)^2\,dq$.
Eq.~(\ref{g}) is the efficient estimator for $g$, which saturates the Fisher information~(\ref{Fisher}) in absence of noise. The inverse of the Fisher information is known as the Cramer-Rao bound or CRB. This bound is the smallest possible variance of an unbiased estimation of the parameter $g$. The smallest possible standard deviation for measurements of $g$ is given by $\Delta g^{CRB}=F_g^{-1/2}=1/\left(2\sqrt{N}\sigma\right)$, and any source of noise would increase it. This result shows the standard quantum limit or shot-noise dependence with respect to $\sqrt{N}$, characteristic of $N$ independent measurements. 



\paragraph{Comparison to other techniques}
We now consider two alternative approaches to measure the parameter $g$, and compare them to our proposed protocol. We note that all three techniques are upper bounded by the same Fisher information, $4N\sigma^2$. 

The first approach is formulated by noticing that the unitary evolution corresponds to a translation of $g$ in the canonical momentum of $q$, and that the ancillary system $\hat{A}$ is not necessary. The protocol consists on preparing $N$ copies of $\psi(q)$, applying the evolution $\hat{U}_{st}=\exp(-ig\hat{q})$, and performing measurements of the conjugate momentum $p$ (instead of $q$). The visibility of the shift, i.e. the ratio of the shift and the standard deviation of the measured state, is given by $\delta p/\sigma_p=g/(1/2\sigma)=2g\sigma$. This approach is known as the \textquotedblleft standard technique\textquotedblright~when similar comparisons to WVA are introduced. The visibility of our technique takes the form $\delta q/\sigma=(2g\sigma^2\cot\epsilon)/\sigma=2g\sigma\cot\epsilon$, giving an advantage of $\sim1/\epsilon$ for $\epsilon\ll1$ when measuring small values of $g$. This amplification is an important key on how the WVA and our protocol are superior in technical-noise-limited experiments. 

The second approach is the conventional WVA technique, where measuring $q$, using only one detector, is conditioned to the postselection $|\Psi_f\rangle=\left(|0\rangle-e^{i\epsilon}|1\rangle\right)/\sqrt{2}$.  This procedure is equivalent to removing the balancing phase $\pi/2$ in Fig.~\ref{fig:cartoon} and tracking only the dark port of the interferometer. The probability distribution for such a measurement is given by $P_2^{\mbox{WVA}}(q)=\sin^2{(\epsilon/2)}\,P(q-2g\sigma^2\cot(\epsilon/2))$. So, even though the shift of the peak of our protocol is half the shift of the peak in the WVA technique for a given $\epsilon$ (for $\epsilon\ll1$), the peak value of $\tilde{P}_-(q)$ is much larger than in $P_2^{\mbox{WVA}}(q)$ since $\epsilon\gg\epsilon^2/4$. This result plus the background noise subtraction characteristic of differencing signals allow us to experimentally induce smaller possible values of $\epsilon$ than in WVA, which offers technically advantageous larger amplification than in WVA.

The Fisher information for the WVA technique, under the weak interaction approximation, is given by $F_g^{\mbox{WVA}}=4N\sigma^2\cos^2(\epsilon/2)$~\cite{Supplemental}. WVA measurements \textit{asymptotically} recover all of the Fisher information, given by $4N\sigma^2$, only in the anomalous weak value regime ($2g\sigma\ll\epsilon/2\ll1$)~\cite{PRX}, but the proposed differencing technique collects always \textit{all} of the information. WVA and the almost-balanced weak values technique offer similar amplification behaviour, and noise mitigation advantages will depend strongly on the metrological specific experimental task. For example, WVA could still be a favourable technique in situations where detector saturation is the predominant limiting factor.   
 

Finally, note that both of the above techniques (standard and WVA) require prior measurements of the unshifted peak and variance of the input distribution $P(q)$. Ours does not, since the sum distribution $\tilde{P}_+(q)$ offers these measurements simultaneously. This is one of the advantages of using two detectors instead of just one.  

\begin{figure}
\centering
\includegraphics[width=0.45\textwidth]{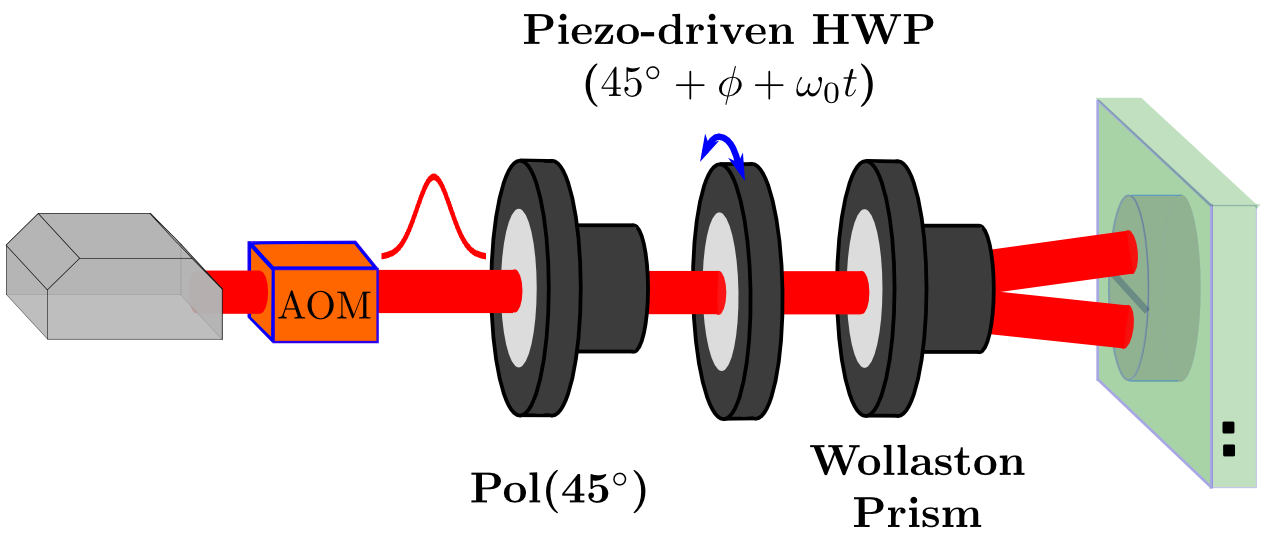}
\caption{Experimental setup for measuring small angular velocities $\omega_0$ of a piezo-driven half wave plate (HWP). The angular rotation induces changes of polarization in the laser field.}
\label{fig:setup2}
\end{figure}

 
\paragraph{Experimental implementation}
As a proof-of-principle of the technique in the optical domain, we performed measurements of small polarization changes in a linearly polarized pulse of laser light. A piezo-driven half-wave plate (HWP) played the role of the interferometer in Fig.~\ref{fig:cartoon}, where we used polarization instead of the which-path degree of freedom, and time as the variable $q$ (with $\sigma=\tau$). The piezo actuator rotated the HWP in time by an angle $\phi+\omega_0t$, which defined the tunable phase as $\epsilon=4\phi$, and the angular velocity of the rotating HWP as the parameter of interest, $g=2\omega_0$ (see Fig.~\ref{fig:setup2} and supplemental material for experimental details).  
%
%
The estimates for $\omega_0$ using a split detector showed an almost perfect linear response and small standard deviations even without the need of pulse averaging. The angle $\phi$ was $4.972(5)$ mrad on average. This unbalanced phase is \textit{significantly smaller} than previously reported postselection angles using WVA techniques, proving the possibility of larger amplification under our proposed protocol. 
%

Besides the two (standard and WVA) experimental techniques mentioned before, we compare here our protocol to a conventional balanced optical technique for measurements of $\omega_0$, where no ancillary system nor measurements of peak shifts in a distribution are required. By replacing the Gaussian pulse with a continuous wave (c.w.) beam, the perfectly balanced ($\phi=0$) signal takes the form $I_-(t)= I_0\sin(\omega_0t)$ and the value $\omega_0$ can be recovered. The measured intensities are very small compared to our proposed technique, since $\omega_0\tau\ll\phi$. Thus, the technical advantages of our technique rely principally on time shift measurements of a Gaussian profile with controllable amplitude ($I_0\sin4\phi$), instead of measuring very small voltage amplitudes as it is the case of the c.w. conventional balanced technique. For example, in order to obtain a visible signal (signal-to-noise ratio slightly larger than one) using a c.w. beam, 
1024 periods of integration time were required.
Our technique gives a better signal-to-noise ratio ($\geq$ 29) even without the need of pulse averaging~\cite{Supplemental}.  

\begin{figure}
\centering
\includegraphics[width=0.4\textwidth]{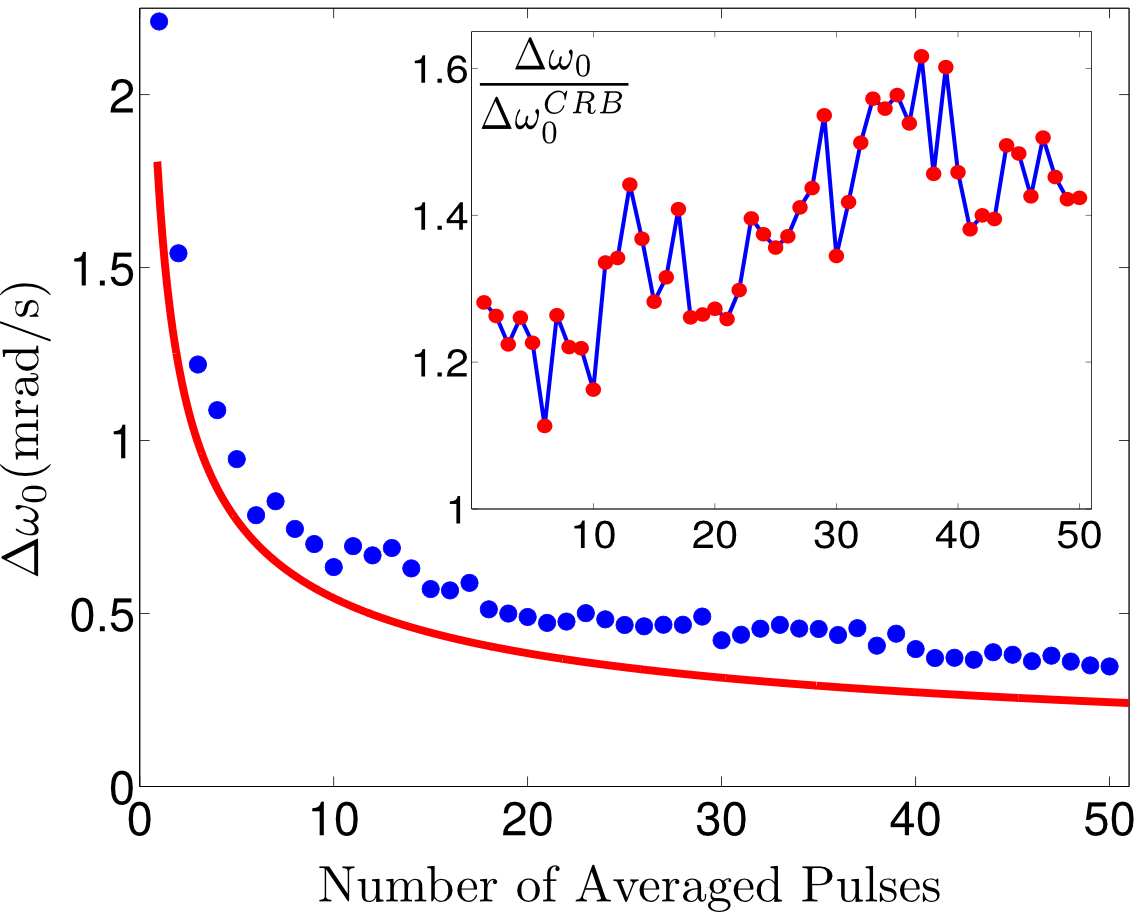}
\caption{Standard deviation on estimations of $\omega_0$ as a function of the number of pulses averaged, $a$. Pulses with $\tau=0.1$ s and $N=5.246(2)\times10^5$ were used. The driving signal on the piezo-mounted HWP was 60 V peak-to-peak at $f_r=1$ Hz. 
Each point (blue) was obtained from the statistics of 100 measurements, and the solid line (red) is given by the shot-noise (eq.~\ref{CRB}). The inset shows the ratio between the experimental standard deviation and the shot-noise.}
\label{fig:PhotonCounting}
\end{figure}

The best possible variance of our measurements are defined by the CRB for $\omega_0$. Following equation~(\ref{Fisher}),
\begin{equation}\label{CRB}
\Delta\omega_0^{CRB}=\frac{1}{\sqrt{F_{\omega_0}}}=\frac{1}{4\sqrt{aN}\tau},
\end{equation}   
where $a$ is the number of averaged pulses. The experimental standard deviations of the measurements with the split detector were estimated to be between 20 and 37 times the calculated ones using Eq.~(\ref{CRB}). Such deviation is small considering that no frequency filters, lock-in amplifiers or any other electronic processing device was used. In order to have good photon number statistics and a clearer raw signal, the split detector of Fig.~\ref{fig:setup2} was replaced with two single photon counting modules (SPCM). The input power was attenuated before the Acoustic Optic Modulator (AOM) so that the peak detected photon rate was slightly smaller than $10^6$ counts/s on each SPCM. Fig.~\ref{fig:PhotonCounting} shows the standard deviation on the estimation of $\omega_0$ as a function of the number of averaged pulses $a$, and its comparison to eq.~(\ref{CRB}). The system follows the $(aN)^{-1/2}$ expected shot-noise behaviour, and it has (inset of Fig.~\ref{fig:PhotonCounting}) a precision close to the shot-noise limit. There is a slight decrease in the relative precision ($\Delta\omega_0/\Delta\omega_0^{CRB}$) with the increase on the number of averaged pulses, which is due to long-term drift~\cite{Supplemental}. Fig.~\ref{fig:PhotonCounting} shows results for a total collection time of up to 5000 sec (50$\times$100 pulses) and a factor of no larger than $\sim$1.6 away from the shot-noise limit, in comparison to a factor of $\sim$1.2 for shorter acquisition times.         

\paragraph{Conclusions}

We have introduced a metrological technique for measuring small interaction parameters based on \textit{almost} balancing both signals of a joint measurement. Even though the protocol does not require the postselection of WVA techniques, anomalous amplification is recovered. A small phase controlling the unbalancing of the difference signal plays a similar role to the postselection angle in WVA. By subtraction of the signals, background noise floor is removed and also a linear response with respect to the small phase  $\epsilon$ is obtained. Such linear response is larger than the quadratic response of WVA techniques, and it is the principal advantage of our technique by allowing larger amplification (smaller $\epsilon$) in technical-noise-limited experiments than in WVA techniques. In addition, no prior knowledge of the peak and the variance of the distribution of the input state is required, 
so systematic error-free estimation of the parameter of interest is possible. We believe our technique will find interesting applications on precision metrology, within and outside optical systems. The use of post-selection required for WVA has been challenging in experimental set-ups outside of optics. Our proposed almost-balanced weak values technique overcomes the necessity of post-selection and might have direct applications in non-optical experiments.

Anomalous amplification in our differencing technique allows us to collect all of the Fisher information that WVA techniques collect only in the optimal asymptotic limit of a large/anomalous weak value. This result proves that anomalous amplification can be obtained from the weak system-meter interaction assumption alone, where strong measurements in the system condition the reading in the meter. Technical advantages of our proposed technique and WVA over standard techniques depend upon such a coupling to an ancillary system.



A recent related work reported experimental advantages of the original proposal of Ref.~\cite{Strubi} with respect to the WVA protocol when estimating ultra-small phases~\cite{ExpJointWM}. 

\begin{acknowledgments} 
The authors acknowledge Andrew Jordan and Justin Dressel for valuable suggestions. J.M.R. thanks Bethany Little for a thorough reading of the manuscript. This work was supported by the Army Research Office, Grant No. W911NF-12-1-0263, and by the National Natural Science Foundation of China Grant No. 11374368 and the China Scholarship Council.
\end{acknowledgments}


%

\pagebreak
\widetext
\begin{center}
\textbf{\large SUPPLEMENTAL MATERIAL}
\end{center}


\section{Calculation of the Fisher Information}
Fisher information is the maximum amount of information that can be extracted about an unknown parameter ($g$ in our case) from a random variable $q$. It is calculated as the variance of the score. The score is a measure of the sensitivity of the probability distribution $P(q;g)$ with respect to the parameter $g$, and it is given by
\begin{equation}
V(q;g)=\frac{\partial}{\partial g}\ln P(q;g).
\end{equation}
In our case, we have two different probability distributions, coming from two detectors,
\begin{equation}
P_i(q;g)=\frac{1}{2\sqrt{2\pi}\sigma}\left[1\mp\sin(\epsilon+2gq)\right]\,e^{-q^2/2\sigma^2},
\end{equation}
where $i=\{1,2\}$. The mean value of events collected by each detector is given by
\begin{equation}
N_i=N\,\int_{-\infty}^{\infty}P_i(g;q)dq=\frac{N}{2}\left(1\mp e^{-2g^2\sigma^2}\sin\epsilon\right)
\end{equation} 
and $N=N_1+N_2$ is the total number of measurements, which (in a lossless system) is equal to the number of prepared events.

Note that the mean of the score vanishes,
\begin{eqnarray}
\langle V_i\rangle&=&\int P_i(q;g)\left[\frac{\partial}{\partial g}\ln P_i(q;g)\right]dq\nonumber\\
&=&\frac{\partial}{\partial g}\left[\int P_i(q;g)\,dq\right]=0,
\end{eqnarray}
so the final expression for the Fisher information of $g$ takes the form
\begin{eqnarray}\label{fisher}
F_g&=&F_1+F_2=N_1\langle V_1^2\rangle+N_2\langle V_2^2\rangle\nonumber\\
&=&4N\sigma^2,
\end{eqnarray}
where
\begin{equation}
F_i=N_i\,\int_{-\infty}^{\infty}\frac{1}{P_i(q;g)}\left[\frac{\partial P_i(q;g)}{\partial g}\right]^2dq=2N\sigma^2\left[1\pm\left(1-4g^2\sigma^2\right)e^{-2g^2\sigma^2}\sin\epsilon\right].
\end{equation}
The result in eq.~(\ref{fisher}) is general for any value of $g$ and $\epsilon$, meaning that it is always possible to recover all of the Fisher information by using both detectors. The shift in the difference signal is obtained only under the weak interaction approximation, $2g\sigma\ll\mbox{min}\left\{1,\tan\epsilon\right\}$. To induce the anomalous amplification is also required the almost-balanced condition, $\epsilon\ll1$. In this sense, our proposed technique lies in the regime of $2g\sigma\ll\epsilon\ll1$. 

Now we compare the result of Eq.(~\ref{fisher}) to the Fisher information when using the WVA technique. The intensity in the dark port under the weak interaction approach is given by
\begin{equation}
P^{WVA}(q;g)=\frac{1}{\sqrt{2\pi}\sigma}\,\sin^2{\left(\frac{\epsilon}{2}+gq\right)}\,e^{-q^2/2\sigma^2},
\end{equation}
and after a straightforward calculation we get the number of detected events and the Fisher information,
\begin{eqnarray}
N^{WVA}&=&\frac{N}{2}\left[1-e^{-2g^2\sigma^2}\cos\epsilon\right]\approx N\sin^2(\epsilon/2)+O[g]^2,\\
F^{WVA}_g&=&2N\sigma^2\left[1+(1-4g^2\sigma^2)e^{-2g^2\sigma^2}\cos\epsilon\right]\approx 4N\sigma^2\cos^2(\epsilon/2)+O[g]^2,
\end{eqnarray} 
respectively.

\section{Experimental Methods}

An external cavity 795 nm wavelength diode laser and an Acoustic Optical Modulator (AOM) are used to prepare pulses with an intensity profile $I(t)=I_0\exp\left(-t^2/2\tau^2\right)$. A Glan-Taylor calcite polarizer prepares the polarization state in $|D\rangle=\left(|H\rangle+|V\rangle\right)/\sqrt{2}$, and a piezo-driven half wave plate (HWP) rotating in time by an angle $\phi+\omega_0t$ away from the $|D\rangle$ axis changes the polarization of the pulse. 
We are interested in estimating the angular velocity $\omega_0$ of the rotating HWP (see Fig. 2 in the main document). 

A Wollaston prism was used to separate the horizontal ($|H\rangle$) and the vertical ($|V\rangle$) polarization components, which were independently measured using a segmented quadrant detector. 
The sum and difference signals take the form
\begin{eqnarray}
I_+(t)&=&I(t)=I_0\,e^{-t^2/2\tau^2},\\
I_-(t)&=&I_0\,\sin(4\phi)\,e^{-(t-\omega_0\tau^2/\phi)^2/2\tau^2},
\end{eqnarray}
assuming that $\omega_0\tau\ll\phi\ll1$. These pulses were recorded and the angle $\phi$ and the time shift $\delta t=\omega_0\tau^2/\phi$ obtained from numerically fitting the distributions. We estimated $\phi$ and $\omega_0$ by using eqs. (5) and (6) of the main document as well, nevertheless  both estimators gave us identical results within the experimental errors.

\begin{figure}[h]
\centering
\includegraphics[width=0.47\textwidth]{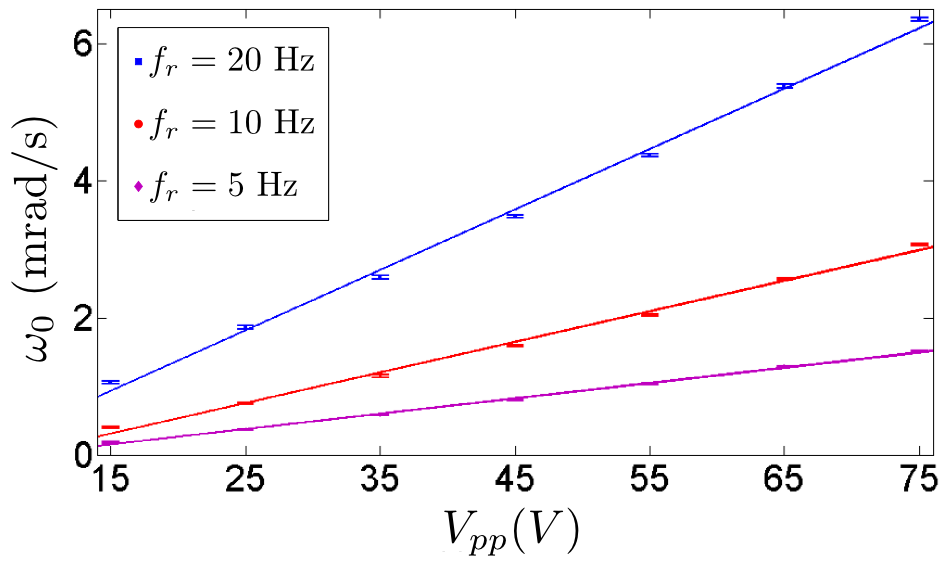}
\caption{Measurements of $\omega_0$ as a function of the peak-to-peak voltage applied to the piezo actuator. The estimated averaged number of collected photons are kept nearly constant, $N\sim2.9\times10^{13}$, by measuring one single pulse when $f_r=$5 Hz, averaging two pulses when $f_r=$10 Hz, and averaging four pulses when $f_r=$20 Hz.
The averaged Signal-to-Noise ratios are between 29 for $V_{pp}=15$ V up to 190 for $V_{pp}=75$ V.}
\label{fig:Results}
\end{figure}

A triangle function in voltage with frequency $f_r$, peak-to-peak voltage $V_{pp}$ and $60\%$ duty cycle as an input for the piezo actuator induced a polarization rotation ramp given by $\omega_0=(10/6)\alpha V_{pp}f_r$, where $\alpha$ is the HWP-mounted piezo response. Gaussian pulses with a peak power of $\sim136\,\mu$W were sent in phase with the ramp. Fig.~\ref{fig:Results} shows the measurement results of $\omega_0$ for three different values of $f_r$ as a function of $V_{pp}$. The technique's estimates show an almost perfect linear response and small error bars even without the need of pulse averaging. The angle $\phi$ was $4.972(5)$ mrad on average. 
The values for the pulse length $\tau$ obtained from numerically fitting the measured sum intensities are 5.377(2) ms for 20 Hz, 10.695(3) ms for 10 Hz, and 21.398(7) ms for 5 Hz. 
In Fig.~\ref{fig:Results} we use the errors propagated from the time shift measurements (error of the mean) which were obtained from the statistics of 60 measured time shifts for each point. The experimental standard deviations of the measurements of Fig.~\ref{fig:Results} were estimated to be between 20 and 37 times the shot-noise (calculated from Eq. (8) in the main document).

We were limited by the resolution of the HWP. We measured $\sim$30 $\mu$rad ($V_{pp}=15$ V) as the smallest reliable total rotation induced by the HWP on the pulses, so we boosted the effective sensitivity of our setup for measuring $\omega_0$ by decreasing the driving frequency on the piezo, $f_r$. Keeping the peak-to-peak voltage on the piezo  at 15 V we set $f_r=2.5$ mHz, $\tau=30.08(1)$ s and $\phi=4.976(6)$ mrad, and measured our lowest reported angular velocity of $\omega_0=161\pm22$ nrad/s. Our reported net rotation of $32\pm4\,\mu$rad in a rotation time of 200 seconds was obtained by averaging sets of five pulses and having 20 statistical points. Our measurements for these long periods were affected by long-term drifts due to temperature and air pressure fluctuations, so 100 consecutive pulses (out of 360 collected) were selected such that the angle $\phi$, the peak laser power and the measured time shift remained relatively constant. 

\section{Polarization/Phase Rotations}
We prepare the system in the state
\begin{equation}\label{initialstate}
|\Phi_0\rangle=|\Psi_0\rangle\otimes E(t)=\frac{1}{\sqrt{2}}\left(|H\rangle+|V\rangle\right)\otimes E_0\,e^{-t^2/4\tau^2},
\end{equation}
where $E_0$ is the peak electric field of the Gaussian pulse. We use a classical description for the long non-Fourier-limited pulses of light used in our experiment, but the theory can also be applied to coherent pulses of light or single photons. After passing through the piezo-driven HWP, the polarization is rotated as
\begin{eqnarray}\label{finalstate}
|\Phi_0\rangle &\rightarrow & \frac{1}{\sqrt{2}}\left[\left\{\cos(2\phi+2\omega_0t)-\sin(2\phi+2\omega_0t)\right\}|H\rangle\right.\nonumber\\&&+\left.\left\{\cos(2\phi+2\omega_0t)+\sin(2\phi+2\omega_0t)\right\}|V\rangle\right]\otimes E_0\,e^{-t^2/4\tau^2},
\end{eqnarray}
and after separating the two polarization components, using a Wollaston prism, the recorded intensities in the split detector are given by,
\begin{eqnarray}
I_1&=&\frac{I_0}{2}\left[\cos(2\phi+2\omega_0t)-\sin(2\phi+2\omega_0t)\right]^2\,e^{-t^2/2\tau^2}\\
&=&\frac{I_0}{2}\left[1-\sin\left(4\phi+4\omega_0t\right)\right]\,e^{-t^2/2\tau^2},\,\mbox{and}\\
I_2&=&\frac{I_0}{2}\left[1+\sin\left(4\phi+4\omega_0t\right)\right]\,e^{-t^2/2\tau^2}.
\end{eqnarray}
Finally, calculating the sum and difference signals we get
\begin{eqnarray}
I_+&=&I_1+I_2=I_0\,e^{-t^2/2\tau^2},\,\mbox{and}\\
I_-&=&I_2-I_1=I_0\,\sin\left(4\phi+4\omega_0t\right)\,e^{-t^2/2\tau^2}\\
&\approx&I_0\sin(4\phi)\,e^{\omega_0t/\phi}\,e^{-t^2/2\tau^2}\\
&\approx&I_0\sin(4\phi)\,e^{-(t-\omega_0\tau^2/\phi)^2/2\tau^2},
\end{eqnarray}
where $\omega_0\tau\ll\phi\ll1$.

Note that if we change the basis in equation~(\ref{initialstate}) from $|H\rangle$ and $|V\rangle$ to circularly polarized states,
\begin{eqnarray}
|R\rangle&=&\frac{1}{\sqrt{2}}(|H\rangle-i|V\rangle),\,\mbox{and}\\|L\rangle&=&\frac{1}{\sqrt{2}}(|H\rangle+i|V\rangle),
\end{eqnarray} 
the initial state takes the following representation (up to a global phase)
\begin{equation}
|\Phi_0\rangle=\frac{1}{\sqrt{2}}\left(|R\rangle-i|L\rangle\right)\otimes E_0\,e^{-t^2/4\tau^2}.
\end{equation}
This representation defines an exact mapping of the theoretical protocol presented in Fig. 1 of the main document, where both states ($|R\rangle$ and $|L\rangle$) are separated into the two different paths of the interferometer, and there is a relative phase of $4(\phi+\omega_0t)$ between both arms. In fact, the state after the light pulse passes trough the piezo-driven HWP takes the form (eq.~[\ref{finalstate}]),
\begin{equation}
|\Phi_0\rangle\rightarrow\frac{1}{\sqrt{2}}\left[e^{i2(\phi+\omega_0t)}|R\rangle-i\,e^{-i2(\phi+\omega_0t)}|L\rangle\right]\otimes E_0\,e^{-t^2/4\tau^2}.
\end{equation}
Under this representation, the left and right circularly polarized components of light split into the two arms of the interferometer. If Fourier-limited pulses were used, the quantity $4\omega_0$ could be understood as a relative laser frequency shift between the two paths.

\end{document}